\documentclass[aps,prl,twocolumn,showpacs]{revtex4}

\usepackage{amsmath}
\usepackage{amssymb}
\usepackage{graphicx}
\usepackage{dcolumn}
\usepackage{bm}

\newcommand{\ga}{\gamma}

\newcommand{\al}{\alpha}
\newcommand{\om}{\omega}
\newcommand{\ka}{\kappa}

\newcommand{\br}{{\bf r}}
\newcommand{\prt}{\partial}
\newcommand{\bu}{{\bf u}}

\begin{document}

\title{
Oblique solitons in the flow of polariton
condensate past an obstacle}
\author{A. M. Kamchatnov}
\email{kamch@isan.troitsk.ru}
\author{S. V. Korneev}
\email{svyatoslav.korneev@gmail.com}

\affiliation{
Institute of Spectroscopy, Russian Academy of Sciences, Troitsk,
Moscow Region, 142190, Russia }

\date{\today}

\begin{abstract}
Formation of oblique solitons by a flow of polariton condensate past an obstacle is
considered. The flow is non-uniform due to a finite life-time of polaritons what
changes drastically the conditions of formation of oblique solitons compared with
the atomic condensate case. The theory shows that the polariton solitons
can be generated by a subsonic flow in agreement with the recent experiment [A. Amo {\it et al.},
Science, {\bf 332,} 1167 (2011)]. Geometric form of oblique solitons and other their
parameters are calculated analytically and analytical results are confirmed by numerical
simulations.
\end{abstract}


\maketitle

{\it 1. Introduction.}
Semiconductor microcavities, in which strong coupling of excitons to photon modes leads to the
formation of microcavity polaritons, have become a focus for the study of polariton condensation
and related physical effects in solid state systems \cite{rev1,rev2}. Polaritons possess an
extremely small effective mass $m_{pol}$ of the order of $10^{-4}$ that of electron which allows for
their condensation at temperatures about a few kelvins and more. All
parameters of the condensate can be tuned with the use of resonant lasers and this advantage of
polariton condensate compared with the atomic case has been exploited in the recent observations of
macroscopic coherence \cite{kasprzak-2006,lai-2007,wertz-2010}, quantized vortices \cite{lagoudakis-2008},
superfluid flow past an obstacle \cite{amo-2009a,amo-2009b,amo-2011,grosso-2011}, and persistent superfluid currents
\cite{sunvitto-2010}.

Theoretically, different regimes of the flow of atomic condensate past an obstacle in two-dimensional
(2D) geometry can be distinguished depending on the Mach number $M=u_0/c_s$ ($u_0$ being the velocity
of the incident uniform condensate and $c_s$ the sound velocity in it): (i) for $0<M<0.43$ the flow is
superfluid and no excitations are generated \cite{FPR-1992,WMCA}; (ii) for $0.43<M<1$ vortices are
generated by the flow what corresponds to the usual mechanism of loss of superfluidity; (iii) for
$1<M<1.44$ the channel of Cherenkov radiation of sound waves opens which leads to formation of the interference
wave pattern (``ship waves'') outside the Mach cone \cite{chcs-2006,gegk-2007,gsk-2008}, vortices are
still generated and they are located inside the Mach cone forming oblique vortex streets; (iv) for $M>1.44$
the oblique vortex streets are transformed into oblique dark solitons \cite{egk-2006} which are
surprisingly stable and this is explained as a transition from absolute instability of dark solitons
to their convective instability \cite{kp-2008,kk-2011,hi-2011}.

The dissipative nature of polaritons changes essentially the properties of wave patterns generated by
the flow of polariton condensate past an obstacle. For example, the amplitude of the Cherenkov wave
pattern decays very fast with growth of the distance from the obstacle even in the pumping regime as it was predicted in
\cite{cc-2005} and observed experimentally in \cite{amo-2009a}. Situation changes even more drastically in the flow
of the polariton condensate outside the pumping region \cite{amo-2011} or in the free flow of the condensate
cloud \cite{grosso-2011}. As was found in \cite{amo-2011}, the stable oblique solitons are generates by a 
subsonic flow with the Mach number $M\simeq0.6$ and the above mentioned region (ii) becomes extremely narrow in
the polariton condensate flow. We note that in this experiment the condensate flow was strongly non-uniform---its
density decreased with the distance from the pumping region because of dissipation effects and the flow velocity
increased due to action of pressure, thus the Mach number $M$ became a function of the space coordinate.
In this Letter we develop the theory of oblique dark solitons in such non-uniform situations and show that the dissipation effects relax the condition for convective instability of oblique solitons.
Thus, for subsonic
incident velocity and strong enough damping, the unfavorable for stability region becomes so narrow that
it can be spanned by the oblique soliton wave and this explains the results of Ref.~\cite{amo-2011}.

{\it 2. Theoretical model.}
Polaritons have a finite life-time, and to maintain their steady-state population a
continuous pumping is required. Above a threshold pumping strength an accumulation of low
energy polaritons is accompanied by a significant increase of spatial coherence that extends over the entire cloud of
polaritons \cite{kasprzak-2006,lai-2007} which, hence, with account of repulsive interaction of polaritons caused by their
exciton component, can be described by the Gross-Pitaevskii (GP) equation for the polariton condensate wave function $\psi$.
Due to dissipation the polaritons disappear from the system, and these losses can be modeled by the effective
term $\prt_t\psi=-\ga\psi$. As a result, outside the pumping region, we arrive at the following generalized GP equation
\begin{equation}\label{3-1}
    i\psi_t+\tfrac12\Delta\psi-|\psi|^2\psi=U(\br)\psi-i\ga\psi,
\end{equation}
(written in standard non-dimensional units) where $U(\br)$ denotes the potential
of the obstacle and $\Delta$ is the 2D ($\br=(x,y)$) Laplace operator. In accordance with the experiment \cite{amo-2011},
we assume that the polariton condensate is created in the region $x<0$ with the density $\rho_0$ and it is put in
motion with the flow velocity $u_0$. Hence, in the region $x>0$ it evolves according to Eq.~(\ref{3-1}) and satisfies
the boundary condition $\left.\psi\right|_{x=0}=\sqrt{\rho_0}\exp(iu_0x)$. Numerical solution of this problem is
illustrated in Fig.~1.
\begin{figure}[bt]
\includegraphics[scale=0.4]{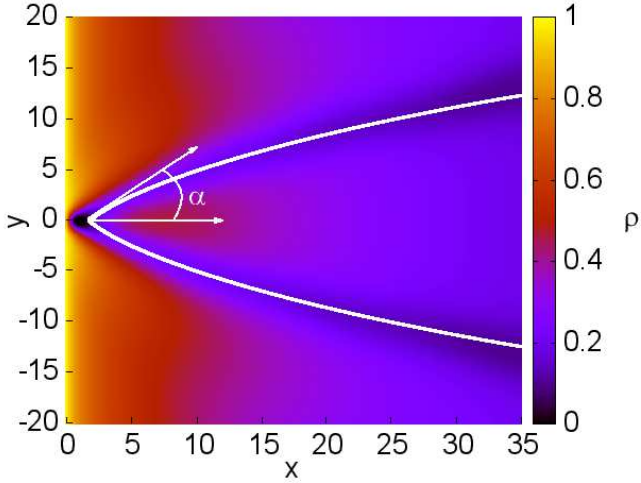}
\caption{(color on line) Distribution of the density plot of polariton condensate in the region $x>0$ for the value
of the damping coefficient $\ga=0.02$. The obstacle is located
at the point $x=1.0,\,y=0$ and modeled by the potential $U$ corresponding to an impenetrable disk with diameter
equal to unity. The parameters of the incident flow are $\rho_0=1.0,\,u_0=0.65$.
White curves show the locations of the
solitons according to the analytical formulae (see (\ref{5-1a}) and (\ref{6-2a})).}
\label{fig1}
\end{figure}
Similar pattern was found in numerical solution of a more general model
\cite{pcc-2011}. However, simpler
model (\ref{3-1}) is enough for our aim to understand the behavior of oblique solitons with account of
dissipation and it permits us to obtain more detailed analytical results, so we will study it here.
As we see in Fig.~1, the distribution of density decays with growth of $x$ and the oblique
solitons are created by a subsonic incident flow. Due to non-uniformity of the flow, the oblique
solitons are curved. Thus, our task is to develop the theory of these effects and to give simple
estimates of the main parameters of the flow and of solitons.

It is convenient to transform Eq.~(\ref{3-1}) to the hydrodynamic form by means of the substitution
\begin{equation}\label{4-1}
    \psi=\sqrt{\rho}\,\exp\left(i\int^{\br} \bu(\br',t)d \br'\right)
\end{equation}
which yields
\begin{equation}\label{4-2}
    \rho_t+\nabla\cdot(\rho u)=-2\ga\rho,
\end{equation}
\begin{equation}\label{4-3}
\bu_t+(\bu\cdot\nabla)\bu+\nabla\rho+\nabla\cdot\left(\frac{(\nabla\rho)^2}{8\rho^2}
-\frac{\Delta\rho}{4\rho}\right)=-\nabla U.
\end{equation}
Then the boundary conditions reduce to
\begin{equation}\label{4-4}
    \rho=\rho_0,\quad u=u_0 \quad \text{at}\quad x=0.
\end{equation}

Equation (\ref{3-1}) with $\ga=0$, $U(\br)\equiv0$ and uniform background density
$\rho=\rho_{b}$ has the dark soliton solution
which in the reference system with the quiescent condensate ($u_0=0$) can be written in the form
\begin{equation}\label{4-5}
\begin{split}
    \rho_s(x,t)&=\rho_{b}\left\{1-\frac{1-V^2/\rho_{b}}{\cosh^2[\sqrt{\rho_{b}-V^2}\,(x-Vt)]}\right\},\\
     u_s(x,t)&=V(1-\rho_{b}/\rho_s(x,t))
\end{split}
\end{equation}
where $x$ axis is the direction of propagation.
The oblique solitons of Ref.~\cite{egk-2006} far enough from the obstacle can be considered as
solutions (\ref{4-5}) transformed to the reference system with the obstacle at rest and such a value of the
velocity $V$ that it is canceled by the component of the flow velocity normal to the soliton.
We have to find how
this solution deforms in a non-uniform condensate.

{\it 3. Hydraulic approximation.} If the dependence of the background steady flow on $x$ coordinate is not
too fast, then it can be found in the hydraulic approximation with neglected dispersion terms in Eq.~(\ref{4-3})
and the time derivatives put equal to zero. The flow becomes smooth enough at some $x=x_1$ where $\rho=\rho_1$
and $u=u_1$. The system can be easily reduced to a single equation
\begin{equation}\label{5-1}
    \left(\rho\sqrt{u_1^2+2(\rho_1-\rho)}\right)_x=-2\ga\rho
\end{equation}
which solution with the initial condition $\rho=\rho_1$ at $x=x_1$ gives the background density $\rho_b(x)$ for $x\geq x_1$.
It is given in an implicit form by the equation
\begin{equation}\label{5-1a}
x=x_1+(\sqrt{\rho_1}/\ga)X(\rho_b/\rho_1),
\end{equation}
where
\begin{equation}\label{5-2}
    \begin{split}
    &X(\rho)=\tfrac32(M_1-[M_1^2+2(1-{\rho})]^{1/2})+(M_1^2+2)^{1/2}\\
    &\times\ln \left[\sqrt{{\rho}}\cdot
\frac{(M_1^2+2)^{1/2}-M_1}{(M_1^2+2)^{1/2}-[M_1^2+2(1-{\rho}]^{1/2}}\right],
    \end{split}
\end{equation}
$M_1=u_1/\sqrt{\rho_1}$ and
\begin{equation}\label{5-3}
u_b(\rho_b)=\sqrt{u_1^2+2(\rho_1-\rho_b)}
\end{equation}
is the background flow velocity. These formulae determine the distributions of the density $\rho_b$ and the
flow velocity $u_b$ as functions of $x$. If the incident flow is supersonic, $u_0>\sqrt{\rho_0}$, then they are applicable
in the entire region $x>0$, so that $x_1$ can be put equal to zero and, hence, $\rho_1=\rho_0$ and $u_1=u_0$.
For a subsonic incident flow velocity $u_0<\sqrt{\rho_0}$ the profiles change very fast near the input plane
$x=0$ and the dispersion effects cannot be neglected. However, at the distance about one healing length
the profiles become smooth enough for applicability of the hydraulic approximation (\ref{5-2}),(\ref{5-3}),
hence we can choose the reference point as $x_1\sim1$.
As we see, the parameters of the flow change considerably at the distance about $\sqrt{\rho_1}/\ga$, that is of
the sound velocity multiplied by the time-life of polaritons.
Accuracy of the formulae obtained is illustrated in Fig.~2 by the
plot of a local Mach number $M(x)=u_b(x)/\sqrt{\rho_b(x)}$ as a function of $x$. As we see, for typical
values of the parameters the Mach number exceed
the critical value $M=1.44$ of transition from absolute instability of oblique solitons 
to their convective instability very close
to the obstacle at the distance much less than the size $l\sim\sqrt{\rho_0}/\ga$
of the condensate and the hydraulic approximation is applicable practically everywhere.
\begin{figure}[bt]
\includegraphics[scale=0.4]{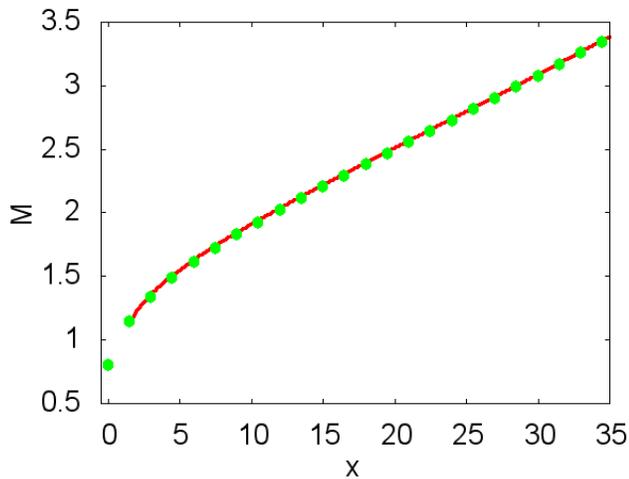}
\caption{(color on line) The local Mach number $M(x)=u_b(x)/\sqrt{\rho_b(x)}$ as a function of $x$. The parameters of the incident flow are $\rho_0=1.0,\,u_0=0.65$; $\ga=0.02$. The continuous line corresponds to the analytical formulae and the dots to the
numerical solution of Eq.~(\ref{3-1}). }
\label{fig2}
\end{figure}

{\it 3. Dark soliton on the slowly dissipating background.}
Motion of dark solitons along non-uniform background, e.g., along the 1D Thomas-Fermi profile formed by the
stationary trap with a harmonic potential, was studied in \cite{ba-2000} and it was
shown in \cite{ba-2001,kp-2004} that dynamics of dark solitons is determined by conservation of the soliton's energy. In our dissipative case
neither the number of polaritons, nor the energy is preserved and the stationary profiles
(\ref{5-2}),(\ref{5-3}) are created by decaying non-uniform flow rather than by the external potential.
We suppose that if the profiles of the density and the flow velocity are varying slowly enough, then
the dark soliton can adjust its local parameters in such a way that its form (\ref{4-5}) is preserved
with varying density $\rho_b=\rho_b(\br)$, velocity $V=V_1\sqrt{\rho_b(\br)/\rho_1}$, and $x-Vt$ replaced by the coordinate
$\xi$ normal to the curve $y=y_s(x)$ of the soliton location (determined, e.g., as a curve along which the
local density changes most slowly). Here $\rho_1$ and $u_1$ are taken at the reference point $\br_1$ which can be
chosen closely enough to the obstacle and then we can assume with good enough accuracy that $V_1=u_1\sin\al$ where
$2\al$ is the opening angle of two oblique solitons generated by the obstacle (see Fig.~1). The curve $y=y_s(x)$
is determined by the condition that the local soliton velocity $V$ is canceled by the normal component of the
local flow velocity, that is
\begin{equation}\label{6-1}
    \frac{dy_s}{dx}=\tan\theta,\quad \sin\theta=\frac{V}u=V_1
    \sqrt{\frac{\rho_b(x)}{\rho_1[u_1^2+2(\rho_1-\rho_b(x))]}}.
\end{equation}
It is easy to see that the solution can be reduced with the use of Eq.~(\ref{5-1a}) 
to a quadrature which in a practically useful limit $V_1\ll\sqrt{\rho_1}$ can be expressed in terms of elementary functions
\begin{equation}\label{6-2a}
    y_s=y_1+(V_1/2\ga)Y(\rho_b/\rho_1)
\end{equation}
where
\begin{equation}\label{6-2}
\begin{split}
    &Y\left({\rho}\right)=3\left(1-\sqrt{{\rho}}\right)+\sqrt{1+M_1^2/2}\\
    &\times\ln\left[\frac{\sqrt{M_1^2/2+1}-1}{\sqrt{M_1^2/2+1}-\sqrt{\rho_b}}
    \cdot\sqrt{1+\frac{2(1-\rho_b)}{M_1^2}}\right].
\end{split}
\end{equation}
Then the functions $x=x_1+(\sqrt{\rho_1}/\ga)X(\rho_b/\rho_1)$, 
$y=\pm [y_1+(V_1/2\ga)Y(\rho_b/\rho_1)]$
determine in parametric form the locations of the dark solitons shown in Fig.~1 by white lines.
Observed agreement of analytical formulae with the numerical results
justifies our supposition that the parameters of oblique soliton adjust their values to
the local parameters of the flow.

{\it 4. Stability of oblique solitons in a non-uniform background.}
As is known, 2D dark solitons propagating along quiescent background are unstable with respect to ``snake''
instability \cite{kp-1970,zakharov-1975,kt-1988} which leads to decay of dark solitons into vortices.
However, dark solitons in BEC can be stabilized by the flow of the condensate which convects the
unstable perturbations away from the obstacle \cite{kp-2008}. As was shown in \cite{kk-2011},  such a
transition from absolute instability of dark solitons to their convective instability corresponds to
the condition that the flow velocity along the soliton exceeds the velocity of propagation of the front of
the instability region and this front velocity equals to the minimal group velocity of the waves
propagating along the soliton. To apply this theory to our problem, we have to generalize it to the
case of dark solitons propagating through a non-uniform background.

The spectrum $\om(p)$ of harmonic waves $\propto\exp[i(p\eta-\om t)]$
propagating along dark solitons (here $\eta$ is a coordinate along the soliton) was calculated in \cite{kt-1988}
where it was shown that for wave numbers $0<p<p_c$, where
\begin{equation}\label{7-1}
p_c(V)=\left[2\sqrt{V^4-V^2+1}-(1+V^2)\right]^{1/2},
\end{equation}
$V$ being velocity of the soliton moving with respect to the quiescent condensate with a uniform
density $\rho_b=1$, the spectrum is pure imaginary, i.e. it corresponds to unstable modes, and for
$p>p_c$ it is real, i.e. it corresponds to propagating modes. The analytical expressions for $\om(p)$
are known in two limits: for $p\ll p_c$ (see \cite{kt-1988}) and for $|p-p_c|\ll p_c$
(see \cite{psk-1995}). We will use here an accurate enough interpolation formula
\begin{equation}\label{7-2}
\begin{split}
\om(p,V)&\cong \sqrt{\frac{1-V^2}{3p_c(V)}}\,p(1+\beta(V)p)\sqrt{p-p_c(V)},\\
\beta(V)&={b(1-V^2)}/{p_c(V)},
\end{split}
\end{equation}
where $b\cong 0.2369$ is a fitting parameter. Then the minimal group velocity equal to the unstable front
velocity is given by the formula
\begin{equation}\label{7-3}
\begin{split}
&v_g(V)={\left[30b\left(\sqrt{1+\tfrac{32}{3}\ka(1+\ka)}-(1+2\ka)\right)\right]^{-1/2}}\\
&\times{\left((1+2\ka)\sqrt{1+\tfrac{32}{3}\ka(1+\ka)}-1+\tfrac{8}{3}\ka(1+\ka)\right)}
\end{split},
\end{equation}
where $\ka=\ka(V)=b(1-V^2)$.

In case of slightly non-uniform background we get the following approximate expression for the
velocity of the instability front as a function of $x$ coordinate of the soliton location point:
\begin{equation}\label{8-1}
\tilde{v}_g(x)=\sqrt{\rho_b(x)}\,v_g(V_1/\sqrt{\rho_1}),
\end{equation}
where the parameters $V_1=u_1\sin\al$ and $\rho_1$ refer to the reference point near the obstacle at which the hydraulic
approximation is applicable with desirable accuracy. If the component of the background flow velocity along the soliton exceeds $\tilde{v}_g(x)$, then at this point $x$ the instability front is convected
away from the obstacle and the oblique soliton increases its length. This condition yields the
inequality
\begin{equation}\label{8-2}
\rho_b(x)\leq\rho_1\cdot\frac{M_1^2+2}{V_1^2/\rho_1+2+v_g^2(V_1/\sqrt{\rho_1})}
\end{equation}
which determines the region of $x$ where the oblique soliton becomes only convectively unstable, i.e.
effectively stable.
The right-hand side of this inequality has a minimum for the most deep solitons with $V=0$ and, hence,
$v_g(0)=1.44$. This gives the upper estimate for the $x_{cr}$ such that (\ref{8-2}) is satisfied
for $x>x_{cr}$: $\rho_b(x_{cr})=0.245\rho_1(M_1^2+2)$, and substitution of this value into
Eq.~(\ref{5-2}) yields the expression for $x_{cr}$,
\begin{equation}\label{8-3}
x_{cr}=\frac{\sqrt{\rho_1}}{\ga}F(M_1),
\end{equation}
where
\begin{equation}\label{9-1}
\begin{split}
&F(M)=\frac32\left(M-\sqrt{0.509(M^2+2)}\right)\\&+
\sqrt{M^2+2}
\ln\left[1.729(\sqrt{M^2+2}-M)\right].
\end{split}
\end{equation}
\begin{figure}[bt]
\includegraphics[width=8.5cm,height=5.5cm,clip]{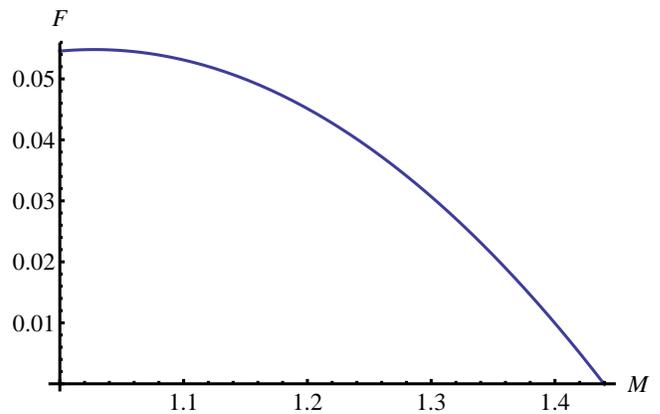}
\caption{Plot of the function $F(M)$ (see Eq.~(\ref{9-1})).}
\label{fig3}
\end{figure}
The plot of the function $F(M)$ is shown in Fig.~3. It vanishes at $M=1.44$, as it
should be, and, what is most remarkable, this function takes very small maximal value
about $F_{max}=0.055$. Just due to this smallness the critical value
$x_{cr}$ is much less than the characteristic size of the decaying condensate
cloud $l\sim \sqrt{\rho_0}/\ga$. For subsonic incident flow the distance
$x_{cr}$ can be calculated numerically only but the result is approximately the
same. Thus, we arrive at the following scenario of formation of oblique solitons
by the flow of polariton condensate. There are two characteristic distances---the condensate
size $l$ and $x_{cr}\sim0.055\cdot l$ differing by more than order of magnitude. If
$x_{cr}$ is about a few healing lengths at the location of the obstacle, then
the shadow ``seeds'' of oblique solitons can easily reach the region of the effective
stability $x>x_{cr}$ and be observed in a wide region  $x_{cr}<x<l$.

Let us make estimates for the parameters of the experiment \cite{amo-2011}.
There the sound velocity at the
location of the obstacle was estimated as $c_s\cong3.5\mu$m/ps and the polariton
life-time as 15ps which gives the condensate's size $l\cong 50\mu$m as was observed
in the experiment. Then we get $x_{cr}\cong 0.055\cdot l\cong 3\mu$m which is about the
observed width of the oblique soliton or the healing length. Thus the oblique solitons
are formed actually in the region of their effective stability what explains their
formation at subsonic Mach numbers of the incident flow.

We are grateful to A.~Amo, N.~Berloff, J.~Bloch,
A.~Bramati, I.~Carusotto, C.~Ciuti, E.~Giacobino, Yu.G.~Gladush, N.~Pavloff, D.~Sanvitto
for discussions of superfluidity in the cavity polariton physics.
This work was supported by RFBR.

\end{document}